\documentclass[sigconf]{acmart}
\def\BibTeX{{\rm B\kern-.05em{\sc i\kern-.025em b}\kern-.08emT\kern-.1667em\lower.7ex\hbox{E}\kern-.125emX}}
    
\copyrightyear{2019}
\acmYear{2019}
\acmConference[]{}{}{}
\setcopyright{rightsretained}

\usepackage[utf8]{inputenc}
\begin{document}

\title{Productization Challenges of Contextual Multi-Armed Bandits}
\author{David Abensur}
\affiliation{%
  \institution{Outbrain Inc}
}
\email{dabensur@outbrain.com}

\author{Ivan Balashov}
\affiliation{%
  \institution{Outbrain Inc}
}
\email{ibalashov@outbrain.com}

\author{Shaked Bar}
\affiliation{%
  \institution{Outbrain Inc}
}
\email{sbar@outbrain.com}

\author{Ronny Lempel}
\affiliation{%
  \institution{Outbrain Inc}
}
\email{rlempel@outbrain.com}

\author{Nurit Moscovici}
\affiliation{%
  \institution{Outbrain Inc}
}
\email{nmoscovici@outbrain.com}

\author{Ilan Orlov}
\affiliation{%
  \institution{Outbrain Inc}
}
\email{iorlov@outbrain.com}

\author{Danny Rosenstein}
\affiliation{%
  \institution{Outbrain Inc}
}
\email{drosenstein@outbrain.com}

\author{Ido Tamir}
\affiliation{%
  \institution{Outbrain Inc}
}
\email{itamir@outbrain.com}
\date{}

\begin{abstract}
Contextual Multi-Armed Bandits is a well-known and accepted online optimization algorithm, that is used in many Web experiences to tailor content or presentation to users' traffic. Much has been published on theoretical guarantees (e.g. regret bounds) of proposed algorithmic variants, but relatively little attention has been devoted to the challenges encountered while productizing contextual bandits schemes in large scale settings. This work enumerates several productization challenges we encountered while leveraging contextual bandits for two concrete use cases at scale. We discuss how to (1) determine the context (engineer the features) that model the bandit arms; (2) sanity check the health of the optimization process; (3) evaluate the process in an offline manner; (4) add potential actions (arms) on the fly to a running process; (5) subject the decision process to constraints; and (6) iteratively improve the online learning algorithm. For each such challenge, we explain the issue, provide our approach, and relate to prior art where applicable. 
\end{abstract}



%
% Keywords. The author(s) should pick words that accurately describe the work being
% presented. Separate the keywords with commas.

\maketitle

\section{Introduction}
Multi-armed bandits is a well known paradigm for managing explor\-ation-exploitation tradeoffs in online learning. Over the past decade, bandits-based schemes have proven successful in optimizing many Web experiences. In particular, Contextual Bandits (CMAB) schemes were shown to be well-suited to online experiences, with context often representing traffic or user characteristics \cite{li2010contextual, li2011unbiased, tang2015personalized, tang2013automatic, Tewari2017}.

In the stochastic contextual bandits model, the world presents a sequence of requests to an algorithm. Each request is accompanied by some context vector of dimension $d$. The algorithm can respond to each request with one of $k$ possible actions (=arm pulls), whose reward distributions are (1) unknown; and (2) depend on the context. After choosing an action, its reward is observed and the algorithm can adapt. Given a context and an action, the rewards are assumed to be independent and identically distributed.

Much of the literature on multi-armed bandits in general and contextual bandits in particular has focused on theoretical guarantees of the various algorithms, such as regret bounds. In contrast, relatively little has been published on the challenges facing the productization and ongoing operational application of contextual bandits schemes in large scale Web services. We address such aspects by discussing (1) tailoring the actual context to be used to the traffic volume of the optimized experience; (2) sanity-testing a running optimization process; (3) performing efficient offline analysis of CMAB schemes; (4) adding actions to already running optimizations; (5) respecting constraints that limit the actions that may be taken at each point in time; and (6) designing the system so as to enable iterative improvement of models. We explain the importance of each challenge and
how we addressed it within a concrete system whose use cases and architecture we describe. Our solutions are practical, and in specific cases -- e.g. the methodologies for monitoring continuity and stability of CMAB processes (Section~\ref{subsec:sanity}) and the extension of the {\em replay} method  (Section~\ref{subsec:offline}) -- constitute novel contributions in and of themselves.

%The rest of 
This paper is organized as follows. Section~\ref{sec:related} surveys related work.
Section~\ref{sec:ref} describes the use cases which we solve with contextual bandits, and the  architecture of our contextual bandits system. Section~\ref{sec:challenges} enumerates the challenges that any product-grade usage of contextual bandits should address, and presents how our system addresses those challenges. We conclude in Section~\ref{sec:conc}.
\section{Related Work}
\label{sec:related}

The contextual bandits setting appears in the literature in many different names and flavours including {\em bandit problems with side observations}~\cite{WANG2005903}, {\em bandit problems with side information}~\cite{lu2010contextual}, and {\em bandit problems with covariates}~\cite{sarkar1991one}. The term {\em contextual multiarmed bandits} was coined by Langford and Zhang~\cite{langford2008epoch}.
%In addition, multi-armed bandit problems can be seen as a subset of reinforcement learning type problems. 
%CMAB algorithms are able to represent the state of the world by mapping onto available arms or actions. Therefore, 
CMAB algorithms have been leveraged in many applications, from recommendation engines and advertising~\cite{lai1985asymptotically, li2010contextual, tang2013automatic, tang2015personalized} to medicine and healthcare~\cite{Tewari2017}. See \cite{burtini2015survey, Tewari2017} for detailed surveys.

Previous work examined how the performance of bandit schemes, which are inherently online, may be accurately evaluated in an offline manner. A commonly used technique is called {\em Replay}~\cite{li2011unbiased, mary2014improving,langford2008replay}. Swaminathan and Joachims framed this as a counterfactual risk minimization problem~\cite{swaminathan2015counterfactual}. 

\section{Use Cases and Architecture}
\label{sec:ref}

\begin{figure}[tb]
\includegraphics[scale=0.32]{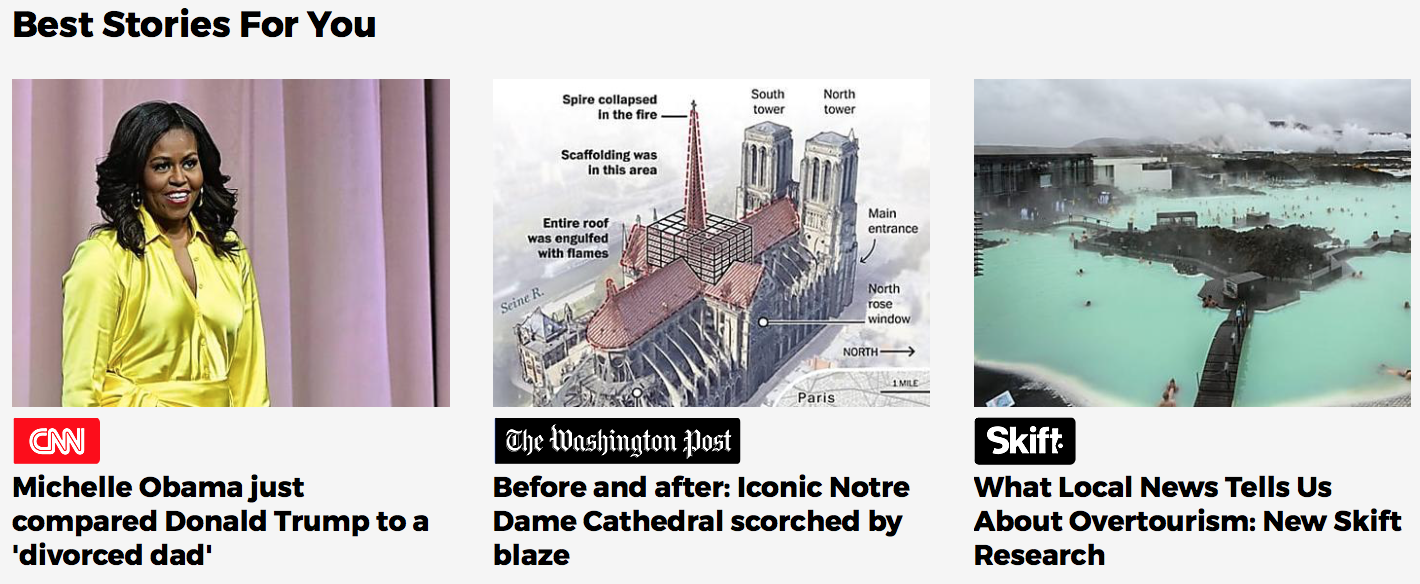}
\centering
\caption{Discovery Widget on the Web}
\label{discovery}
\end{figure}

\begin{figure}[t]
\includegraphics[scale=0.41]{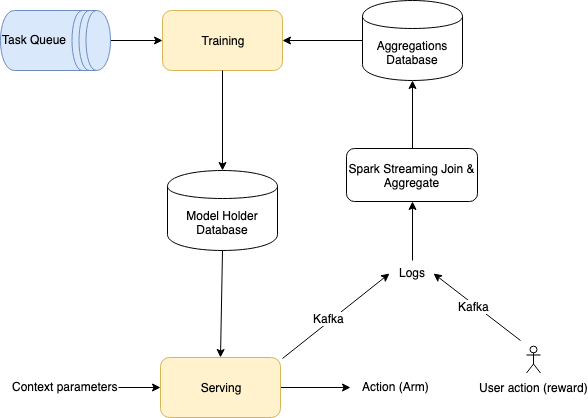}
\centering
\caption{Reference Architecture of CMAB Serving}
\label{fig:arch}
\end{figure}

We apply CMAB schemes -- LinUCB~\cite{li2010contextual} in particular-- with a unified learning and serving architecture, to two business problems:
\begin{description}
\item[UI Optimization:] we serve billions of discovery widgets each day (see Figure~\ref{discovery}). We have observed that seemingly small changes in the appearance or rendering of the widgets may dramatically impact user engagement. We thus have multiple designs of widget styles, modeled as bandit arms, that are selected and served given the context of the request (e.g. device type and screen size). 
\item[Feed Optimization:] in addition to serving standalone widgets, we also support {\em discovery feeds}. A discovery feed is an infinite scroll experience that serves additional recommendations as the user scrolls through previous ones. In practice, our feeds are composed of a sequence of {\em typed cards}, where each card type is a coherent set of recommendations (e.g. about some common theme, or belonging to a certain vertical). The optimization problem here is to select the next card type to serve (those are the arms), given the context of the request and the last few cards already served in the feed. In essence, we are solving for {\em order and frequency} of card types. 
\end{description}
To satisfy both these needs, we designed a system -- depicted in Figure~\ref{fig:arch} -- comprised of two layers, an offline training layer using aggregated data and and online serving layer.\footnote{A similar architecture was proposed by the FAME system~\cite{lempel2012hierarchical}, which was also designed to optimize, among others, rendering and layout use-cases.}

%Before diving into the details of the architecture, we need to clarify the notion of context. In this paper, a context is represented by multiple features such as the \textit{os}, \textit{platform}, \textit{time of day},... This defines the granularity over which we want our models to learn.

The offline layer is comprised of an {\em Aggregations Database}, a {\em Training Service}, and a {\em Task Queue}.
The Task Queue is aware of all the different CMAB-instances running in the system (all optimization use cases), and enqueues periodically (every couple of minutes) requests to the Training Service to update the model per each active instance. Each request contains the set of actions (arms) that are active in that instance.

The Training Service holds in memory the models of all active CMAB-instances. It pulls update requests from the Task Queue, and updates the relevant model, namely the weights per each arm of that model. It does so by reading tuples stored in the Aggregations Database. Unlike theoretical bandit models, which are sequential decision processes where the model makes one decision at a time, immediately observing its reward and updating itself, the Web reality is different. Our serving layer makes thousands of decisions per second. Rewards, mostly in the form of user clicks, arrive asynchronously several minutes after our serving decision has been made, and in particular after the model may have been called upon to perform tens or hundreds of thousands of subsequent decisions. We thus aggregate decisions and rewards in mini-batches, spanning a couple of minutes of accumulated data, where each mini-batch includes tuples of  $\{\mbox{context, arm, number of pulls, observed reward}\}$: the number of times an arm was pulled in a context, and the overall reward resulting from such pulls.

Once the Training Service updates a model of a CMAB-instance, it stores the output in the {\em Model Holder Database}, which acts as a data interface between the offline and serving layers.

Moving to the serving layer, it handles requests that correspond to active CMAB-instances. Per request, it computes in real-time scores for all available arms and returns the scores. The decision or action corresponding to the highest-scoring arm is then served, and both context and decision are logged. Subsequent user interactions such as clicks will also be logged with that same metadata, and the join of serving decisions and resulting user interactions is aggregated into the Aggregations Database to be leveraged in the next batch of model updates.

\section{Challenges}
\label{sec:challenges}
\subsection{Determining Context}
Contextual Bandits literature assumes that the context of each arm pull is well defined. One of the biggest challenges in applications of CMABs is determining the context vector to apply to actual requests. There is often some broad world context $C^w$ which needs to be projected into a simpler context $C$ that will actually be plugged into the model. That projection is essentially a form of feature engineering that should consider the interplay between the context space and the amount of traffic (arm pulls and rewards) that the model will face. Optimizing a small-traffic situation while using a large feature space is a recipe for over-fitting, especially when using popular yet simple models such as LinUCB~\cite{li2010contextual}, that do not inherently reduce dimensionality as part of their internal operation.

In our setting -- prior to starting the incremental learning of our CMAB model -- no reward data is available for guiding feature selection and engineering, nor for tuning hyper parameters.
Since our (full) projected context $C$ might itself be too sparse to enable effective generalization in cases of limited traffic, we further project it into a coarser grained context $C'$, where each context dimension $C_i'$ is a binned representation of the original corresponding contextual dimension $C_i$.
We then plug the unified context $C^u=C \cup C'$, into our CMAB model, enabling early generalization based on $C'$ followed by further refinement, based on $C$, as additional arms are pulled and rewards are observed. 

To further reduce overfitting due to a large %attained 
context relative to limited traffic data, we employ regularization. In the context of the LinUCB algorithm,  regularization is introduced by a  parameter $\lambda$ multiplying the identity matrix used in the Ridge Regression formula. 
As $\lambda$ cannot be tuned in advance before model initiation, we set it to some initial value that is later periodically adjusted based on replays of the model on recently logged data (See Section~\ref{subsec:offline}).

Lastly, different forms of unsupervised dimensionality reduction may be employed such as PCA or Random Projection \cite{Yu2017CBRAPCB}. Such methods may also require matching the reduced context cardinality to the amounts of available traffic, but are inherently less susceptible to a potential "curse of dimensionality".

%Another employed method is maintaining an ensemble of two models $M$ and $M'$ based on the corresponding contexts $D$ and $D'$. Our %final prediction is a weighted average of both models, defined as $\alpha \cdot Pred_M+(1-\alpha) \cdot Pred_{M'} $ where %$0<\alpha<1$ is being constantly updated based on the performance of both models within a given sliding time window. The disadvantage %of this approach is that both models are able to leverage only half of the available data for their learning.

\subsection{Sanity Testing a Running Process}
\label{subsec:sanity}
Imagine a CMAB process that has been running for a while. Especially if the context is rich, it may be difficult to determine whether the process is converging to a state that is "making sense", or whether there is some over-fitting or instability in its results.
Two tests which we use to validate whether the algorithm has picked up some signal are (1) checking continuity; and (2) checking stability.

When checking for continuity, the basic assumption is that if the model outputs a distribution $D_c$ over the arms when given context $c$, the distribution $D_{c'}$ should be similar to $D_c$ whenever $\vert\vert c-c' \vert\vert$ is small.
When checking for stability, the basic assumption is that if the CMAB outputs a distribution $D_c(t)$ over the arms when given context $c$ at time t, then the distribution  $D_c(t')$ should be similar to $D_c(t)$ whenever $\vert t-t'\vert$ is small.

To illustrate how one might test for continuity, we took a use case whose context is defined by a one-hot encoded vector of length 9. We defined the distance between contexts as the Hamming Distance between the vectors. The non-zero distances range from $1$ to  $9$. 
We then used \textit{KL-Divergence} to measure  the distance between the distributions of arms pulled by the algorithm in each context. 

\begin{figure}[h!]
\includegraphics[scale=0.32]{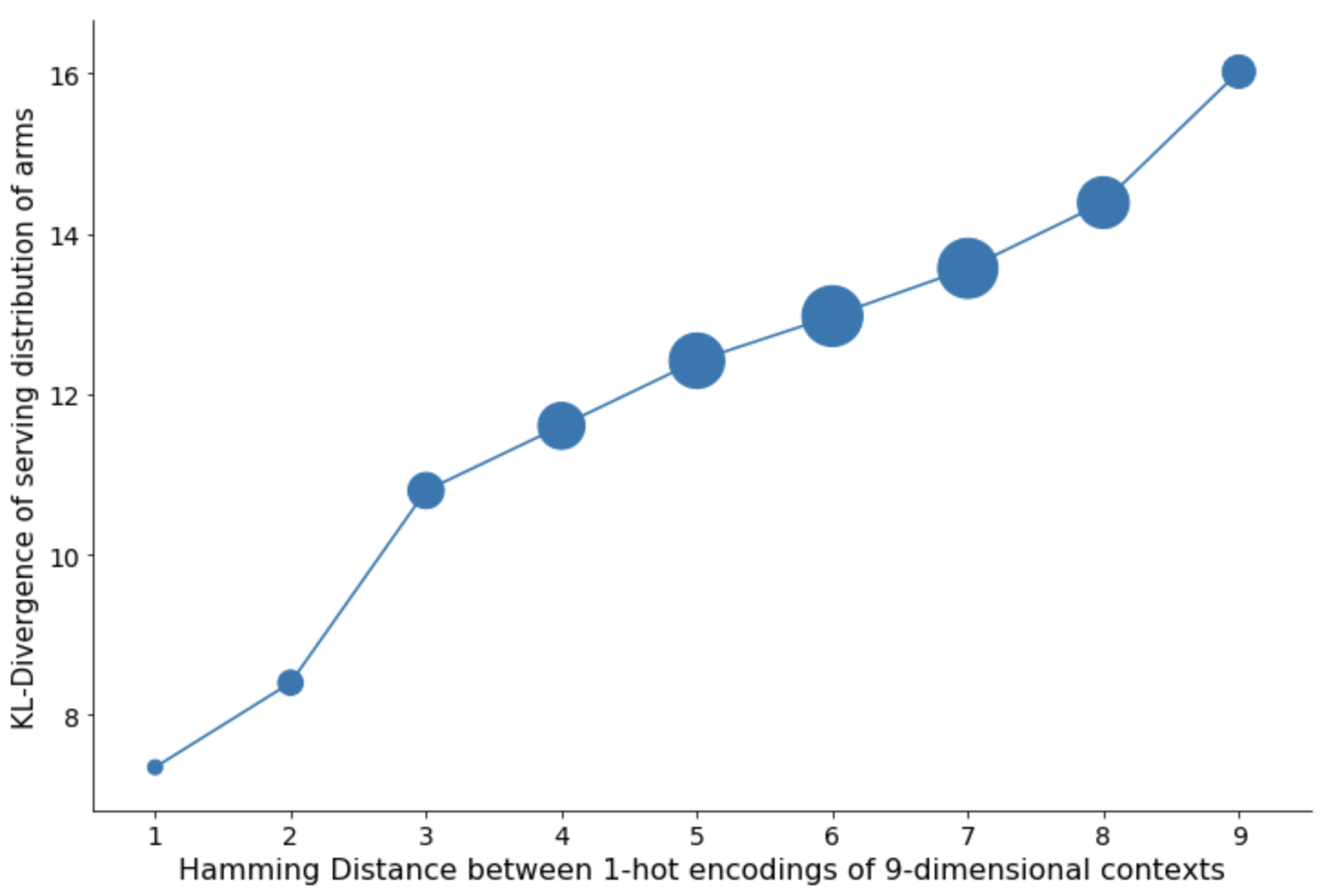}
\centering
\caption{Average KL-Divergence of serving distributions over arms per Hamming Distance between contexts}
\label{continuity}
\end{figure}

Figure~\ref{continuity} presents the average KL-Divergence value for each value of Hamming Distance between contexts. The thickness of each point represents the amount of observed context pairs having the given Hamming Distance.
As expected, we observe that the closer two contexts are, the closer the distributions over the served arms are.
%On top of this,  Here, as we could expect, most of the context pairs are concentrated towards above $50\%$. 

\begin{figure}[h!]
\includegraphics[scale=0.32]{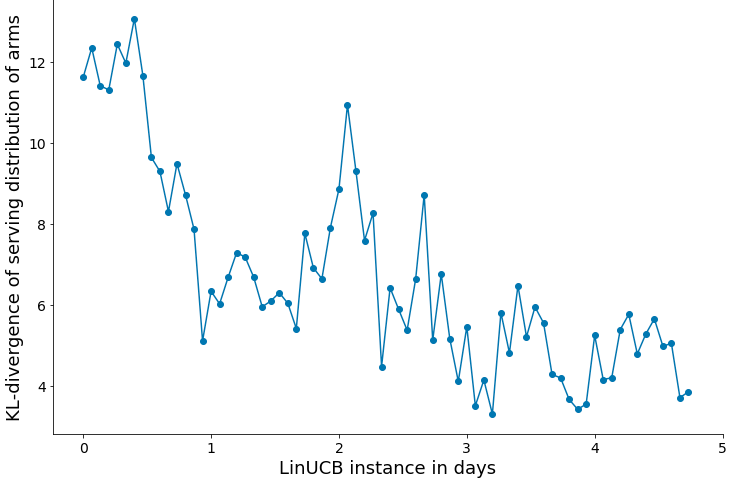}
\centering
\caption{Average difference of serving distributions per time interval of serving of a specific context}
\label{stability}
\end{figure}

Regarding the evaluation of stability, denote by $D_c(t,t+\epsilon)$ the distribution of arms served (pulled) in the time span $(t,t+\epsilon)$ given context $c$. Figure~\ref{stability} plots the average, over all contexts, of 
\[KL(\ D_c(t,t+\epsilon),\  D_c(t+\delta,t+\delta+\epsilon)\ )\] as a function of $t$, the age of the learning instance. We used $\epsilon=10$ minutes and $\delta=1$ hour. We can see that when the instance is young and mostly exploring, one hour of observing rewards can result in large changes in how arms are pulled. As the instance matures, it shifts towards more exploitation, and the hourly changes in the distribution of arm pulls per context become smaller.

%computed the average over all the contexts of the difference of serving distribution of arms for a given context $KL(D_c(t), D_c(t'))$, in function of the time interval between two servings of this same context $\| t - t' \|$. As shown in figure \ref{stability}, we have a clear increase of the distances of distributions below 3600 seconds (e.g. 1h) time difference, which means that every update we perform has high influence of the variants being served. After a while, this increase stabilizes towards the double of the lowest value, so we can assume that - as long as we are in an exploration phase - the distributions of the variants at time $t$ and $t+1h$ are very different. In an exploitation phase however, we should expect the distributions to be much closer.

% Ronny, regarding what is written here relative to exploration and exploitation. I claim that the graph we see represents exploration (from 3.0 to 6.0 for KL div), but I'm not sure this is the case. As most of the tests have been running for a long time now, they must be in an exploitation phase. My guess is that if we tried with real exploration the range of KL divs would be much higher (3.0 to 20.0 for instance), but I don't have room for comparison between both in the paper nor proper time to implement it. What do you think?

\begin{figure}[h!]
\includegraphics[scale=0.32]{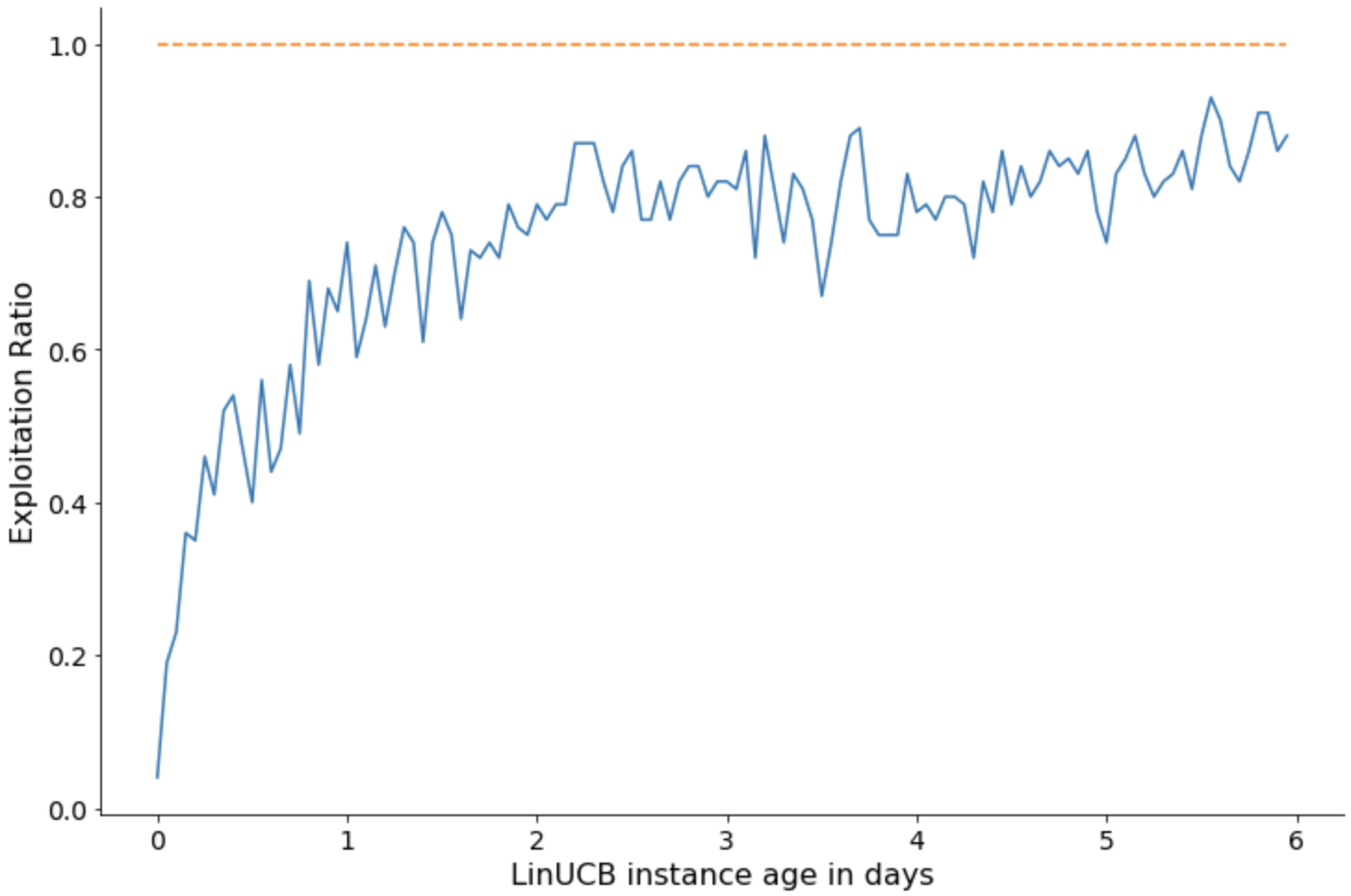}
\centering
\caption{LinUCB Exploitation Ratio as a function of time}
\label{ratio}
\end{figure}

When employing LinUCB, another way to assess its stability is to examine its {\em exploitation ratio}. We define this as the fraction of arms, pulled by the instance, that would have been pulled by a greedy scheme that selects the arm with the highest expected reward, ignoring confidence interval (standard deviation) considerations. Figure~\ref{ratio} plots the exploitation ratio as a function of the age of the LinUCB instance. As Figure \ref{ratio} shows, initially there is little agreement between the actual arms pulled and the greedy leader, which means that the instance is exploring and that  upper confidence bound considerations heavily influence the choice of the arm to pull. Conversely, after 2 days, the instance stabilizes and its arm pulls mostly agree with the greedy choice that maximizes the expected reward, i.e. the instance is mostly exploiting.

\subsection{Offline Analysis}
\label{subsec:offline}
Many machine learned models are trained offline and are measured by offline loss functions. Models exhibiting high loss are shelved. The more promising models can be pushed to production gradually, starting slow (affecting small percentages of traffic) and accelerating as online measurements arrive and show that the new model is performing well. However, this true and tested methodology doesn't translate well to online algorithms such as CMAB processes. In particular, it is counter-productive to deploy such models to production on a small fraction of traffic, since that effectively limits the algorithm's learning. 

To be able to evaluate offline the performance of CMAB models, Langford et al. proposed the \textit{replay} approach \cite{langford2008replay}. To perform  \textit{replay}, one must serve an unbiased portion of traffic by pulling arms uniformly at random. Then, given a CMAB model to evaluate offline, it is fed the same stream of requests served randomly, and a learning step is performed whenever the algorithm selects the same arm that was randomly pulled, using the reward that was observed. This happens with a probability of $1/k$, when $k$ is the number of available arms. The evaluated model's metrics are measured on the rewards of the subset of matching arm pulls. 

The main drawback of the \textit{replay} approach is that it requires serving a lot of random traffic, especially when the number of arms is high. To address this issue, we modified \textit{Replay} as follows. If the evaluated algorithm decides to pull arm $a$ given context $c$ at time $t$, we sample an observed reward from the set of random pulls of arm $a$ given context $c$ in $(t-t_1, t+t_2)$ where $t_1, t_2$ are parameters. In a stationary stochastic setting, $t_1$ and $t_2$ can be set to infinity. In most practical situations, they should be set to some finite application-dependent values, where the distribution of rewards in $(t-t_1, t+t_2)$ is believed to model the reward that would have been observed had arm $a$ been pulled at time $t$. The sampling can be with or without repetitions, where the case of no repetitions sampling might sometimes "run out" of rewards and will not be able to leverage time $t$.
\subsection{Adding Arms on the Fly}
In many cases, a long-running CMAB process has matured, and is serving with little regret per context. The product then admits a new potential action/decision, modeled by the introduction of a new arm. While one can always stop the current CMAB process and start afresh with an expanded set of arms, that is highly inefficient as it loses the accrued learnings of the current model. It is preferable to dynamically add arms on the fly to a running process, continuing to leverage its historical learnings.

We address this challenge by leveraging the Task Queue. It fetches the list of available arms from an external service and passes them to the Training service. Upon encountering a new arm, the training layer will apply the initial model of its learning algorithm (e.g. LinUCB) and will save all arms' models to the Model Holder. In the serving layer, the new arm will compete with the other arms.

\subsection{Subjecting Decisions to Constraints}
In real-life use-cases, some decisions (i.e. arms) may be forbidden in some points in time due to business rules. For example, in our Feed Optimization use-case, it may be forbidden to serve $j$ consecutive cards of the same type in the feed, or we may be required to serve at least one instance of a certain card type within the first $n$ cards. 

Kleinberg et al.'s Sleeping Bandits model \cite{kleinberg2010regret} addresses Stochastic MAB settings when some arms may not be available in certain times. There, the authors proposed the {\em Awake Upper Estimated Reward (AUER)} algorithm and proved its effectiveness. We follow the same intuition in our contextual bandits settings - considering the constraints, the serving layer selects the highest scoring {\em eligible arm} as its action.
\subsection{Iteratively Improving the Model}
We have a certain CMAB model in production, and have an idea for an improvement we believe would drive our metrics even further. Normally, we would run a controlled experiment, a.k.a. A/B test, pitting the production version (control) against the new version (treatment). In a traditional controlled experiment, we would split traffic in an unbiased manner between the control and treatment, annotate in logs which variant served each request, and report on the metrics resulting from each variant. However, that will not suffice for online learning algorithms. Each CMAB process must get its reward only from those requests that it served, and should not have access to results of decisions made by the other variant. Neglecting to do so runs the risk of vicarious reinforcement, wherein variants learn based on the arm pulls of its competitors.

We address this need by logging the test-id and variant-id of any controlled experiment on both its arm pulls and rewards. Our real-time aggregations are subsequently grouped by test and variant ids, and are stored in the Aggregations Database with the test and variant ids being part of the keyspace. Each model of the test reads aggregations from its corresponding keyspace, thus exposing to each variant only its own rewards. 
\section{Conclusions and Future Work}
\label{sec:conc}
This paper presented several underexplored challenges that must be addressed when productizing Contextual Multi-Armed Bandits schemes. While there is an abundance of literature on theoretical aspects of CMABs and on their performance in practice, little has been written about what it takes to promote, monitor, augment and improve such models in a high-scale production environment. This paper covered six such topics, presenting practical solutions to each. In particular, our extension of the {\em replay} method and the methodologies for monitoring continuity and stability of CMAB processes are, to the best of our knowledge, novel in and of themselves.
%\begin{acks}
%If we need non-author acks
%\end{acks}

%
% The next two lines define the bibliography style to be used, and the bibliography file.
\bibliographystyle{ACM-Reference-Format}
\bibliography{productizing-cmab}


\begin{thebibliography}{17}

%%% the Web address.

\ifx \showCODEN    \undefined \def \showCODEN     #1{\unskip}     \fi
\ifx \showDOI      \undefined \def \showDOI       #1{#1}\fi
\ifx \showISBNx    \undefined \def \showISBNx     #1{\unskip}     \fi
\ifx \showISBNxiii \undefined \def \showISBNxiii  #1{\unskip}     \fi
\ifx \showISSN     \undefined \def \showISSN      #1{\unskip}     \fi
\ifx \showLCCN     \undefined \def \showLCCN      #1{\unskip}     \fi
\ifx \shownote     \undefined \def \shownote      #1{#1}          \fi
\ifx \showarticletitle \undefined \def \showarticletitle #1{#1}   \fi
\ifx \showURL      \undefined \def \showURL       {\relax}        \fi
% invisible to TeX
\providecommand\bibfield[2]{#2}
\providecommand\bibinfo[2]{#2}
\providecommand\natexlab[1]{#1}
\providecommand\showeprint[2][]{arXiv:#2}

\bibitem[\protect\citeauthoryear{??}{WAN}{2005}]%
        {WANG2005903}
 \bibinfo{year}{2005}\natexlab{}.
\newblock \showarticletitle{Arbitrary side observations in bandit problems}.
\newblock \bibinfo{journal}{\emph{Advances in Applied Mathematics}}
  \bibinfo{volume}{34}, \bibinfo{number}{4} (\bibinfo{year}{2005}),
  \bibinfo{pages}{903 -- 938}.
\newblock
\showISSN{0196-8858}
\urldef\tempurl%
\url{https://doi.org/10.1016/j.aam.2004.10.004}
\showDOI{\tempurl}
\newblock
\shownote{Special Issue Dedicated to Dr. David P. Robbins.}


\bibitem[\protect\citeauthoryear{Burtini, Loeppky, and Lawrence}{Burtini
  et~al\mbox{.}}{2015}]%
        {burtini2015survey}
\bibfield{author}{\bibinfo{person}{Giuseppe Burtini}, \bibinfo{person}{Jason
  Loeppky}, {and} \bibinfo{person}{Ramon Lawrence}.}
  \bibinfo{year}{2015}\natexlab{}.
\newblock \showarticletitle{A survey of online experiment design with the
  stochastic multi-armed bandit}.
\newblock \bibinfo{journal}{\emph{arXiv preprint arXiv:1510.00757}}
  (\bibinfo{year}{2015}).
\newblock


\bibitem[\protect\citeauthoryear{Kleinberg, Niculescu-Mizil, and
  Sharma}{Kleinberg et~al\mbox{.}}{2010}]%
        {kleinberg2010regret}
\bibfield{author}{\bibinfo{person}{Robert Kleinberg},
  \bibinfo{person}{Alexandru Niculescu-Mizil}, {and} \bibinfo{person}{Yogeshwer
  Sharma}.} \bibinfo{year}{2010}\natexlab{}.
\newblock \showarticletitle{Regret bounds for sleeping experts and bandits}.
\newblock \bibinfo{journal}{\emph{Machine learning}} \bibinfo{volume}{80},
  \bibinfo{number}{2-3} (\bibinfo{year}{2010}), \bibinfo{pages}{245--272}.
\newblock


\bibitem[\protect\citeauthoryear{Lai and Robbins}{Lai and Robbins}{1985}]%
        {lai1985asymptotically}
\bibfield{author}{\bibinfo{person}{Tze~Leung Lai} {and}
  \bibinfo{person}{Herbert Robbins}.} \bibinfo{year}{1985}\natexlab{}.
\newblock \showarticletitle{Asymptotically efficient adaptive allocation
  rules}.
\newblock \bibinfo{journal}{\emph{Advances in applied mathematics}}
  \bibinfo{volume}{6}, \bibinfo{number}{1} (\bibinfo{year}{1985}),
  \bibinfo{pages}{4--22}.
\newblock


\bibitem[\protect\citeauthoryear{Langford, Strehl, and Wortman}{Langford
  et~al\mbox{.}}{2009}]%
        {langford2008replay}
\bibfield{author}{\bibinfo{person}{John Langford}, \bibinfo{person}{Alexander
  Strehl}, {and} \bibinfo{person}{Jennifer Wortman}.}
  \bibinfo{year}{2009}\natexlab{}.
\newblock \showarticletitle{Exploration Scavenging}. In
  \bibinfo{booktitle}{\emph{Proceedings of the 25th international conference on
  Machine learning}}. ICML, \bibinfo{pages}{528--535}.
\newblock


\bibitem[\protect\citeauthoryear{Langford and Zhang}{Langford and
  Zhang}{2008}]%
        {langford2008epoch}
\bibfield{author}{\bibinfo{person}{John Langford} {and} \bibinfo{person}{Tong
  Zhang}.} \bibinfo{year}{2008}\natexlab{}.
\newblock \showarticletitle{The epoch-greedy algorithm for multi-armed bandits
  with side information}. In \bibinfo{booktitle}{\emph{Advances in neural
  information processing systems}}. \bibinfo{pages}{817--824}.
\newblock


\bibitem[\protect\citeauthoryear{Lempel, Barenboim, Bortnikov, Golbandi,
  Kagian, Katzir, Makabee, Roy, and Somekh}{Lempel et~al\mbox{.}}{2012}]%
        {lempel2012hierarchical}
\bibfield{author}{\bibinfo{person}{Ronny Lempel}, \bibinfo{person}{Ronen
  Barenboim}, \bibinfo{person}{Edward Bortnikov}, \bibinfo{person}{Nadav
  Golbandi}, \bibinfo{person}{Amit Kagian}, \bibinfo{person}{Liran Katzir},
  \bibinfo{person}{Hayim Makabee}, \bibinfo{person}{Scott Roy}, {and}
  \bibinfo{person}{Oren Somekh}.} \bibinfo{year}{2012}\natexlab{}.
\newblock \showarticletitle{Hierarchical composable optimization of web pages}.
  In \bibinfo{booktitle}{\emph{Proceedings of the 21st International Conference
  on World Wide Web}}. ACM, \bibinfo{pages}{53--62}.
\newblock


\bibitem[\protect\citeauthoryear{Li, Chu, Langford, and Schapire}{Li
  et~al\mbox{.}}{2010}]%
        {li2010contextual}
\bibfield{author}{\bibinfo{person}{Lihong Li}, \bibinfo{person}{Wei Chu},
  \bibinfo{person}{John Langford}, {and} \bibinfo{person}{Robert~E Schapire}.}
  \bibinfo{year}{2010}\natexlab{}.
\newblock \showarticletitle{A contextual-bandit approach to personalized news
  article recommendation}. In \bibinfo{booktitle}{\emph{Proceedings of the 19th
  international conference on World wide web}}. ACM, \bibinfo{pages}{661--670}.
\newblock


\bibitem[\protect\citeauthoryear{Li, Chu, Langford, and Wang}{Li
  et~al\mbox{.}}{2011}]%
        {li2011unbiased}
\bibfield{author}{\bibinfo{person}{Lihong Li}, \bibinfo{person}{Wei Chu},
  \bibinfo{person}{John Langford}, {and} \bibinfo{person}{Xuanhui Wang}.}
  \bibinfo{year}{2011}\natexlab{}.
\newblock \showarticletitle{Unbiased offline evaluation of
  contextual-bandit-based news article recommendation algorithms}. In
  \bibinfo{booktitle}{\emph{Proceedings of the fourth ACM international
  conference on Web search and data mining}}. ACM, \bibinfo{pages}{297--306}.
\newblock


\bibitem[\protect\citeauthoryear{Lu, P{\'a}l, and P{\'a}l}{Lu
  et~al\mbox{.}}{2010}]%
        {lu2010contextual}
\bibfield{author}{\bibinfo{person}{Tyler Lu}, \bibinfo{person}{D{\'a}vid
  P{\'a}l}, {and} \bibinfo{person}{Martin P{\'a}l}.}
  \bibinfo{year}{2010}\natexlab{}.
\newblock \showarticletitle{Contextual multi-armed bandits}. In
  \bibinfo{booktitle}{\emph{Proceedings of the Thirteenth international
  conference on Artificial Intelligence and Statistics}}.
  \bibinfo{pages}{485--492}.
\newblock


\bibitem[\protect\citeauthoryear{Mary, Preux, and Nicol}{Mary
  et~al\mbox{.}}{2014}]%
        {mary2014improving}
\bibfield{author}{\bibinfo{person}{J{\'e}r{\'e}mie Mary},
  \bibinfo{person}{Philippe Preux}, {and} \bibinfo{person}{Olivier Nicol}.}
  \bibinfo{year}{2014}\natexlab{}.
\newblock \showarticletitle{Improving offline evaluation of contextual bandit
  algorithms via bootstrapping techniques}. In
  \bibinfo{booktitle}{\emph{International Conference on Machine Learning}}.
  \bibinfo{pages}{172--180}.
\newblock


\bibitem[\protect\citeauthoryear{Sarkar et~al\mbox{.}}{Sarkar
  et~al\mbox{.}}{1991}]%
        {sarkar1991one}
\bibfield{author}{\bibinfo{person}{Jyotirmoy Sarkar} {et~al\mbox{.}}}
  \bibinfo{year}{1991}\natexlab{}.
\newblock \showarticletitle{One-armed bandit problems with covariates}.
\newblock \bibinfo{journal}{\emph{The Annals of Statistics}}
  \bibinfo{volume}{19}, \bibinfo{number}{4} (\bibinfo{year}{1991}),
  \bibinfo{pages}{1978--2002}.
\newblock


\bibitem[\protect\citeauthoryear{Swaminathan and Joachims}{Swaminathan and
  Joachims}{2015}]%
        {swaminathan2015counterfactual}
\bibfield{author}{\bibinfo{person}{Adith Swaminathan} {and}
  \bibinfo{person}{Thorsten Joachims}.} \bibinfo{year}{2015}\natexlab{}.
\newblock \showarticletitle{Counterfactual risk minimization: Learning from
  logged bandit feedback}. In \bibinfo{booktitle}{\emph{International
  Conference on Machine Learning}}. \bibinfo{pages}{814--823}.
\newblock


\bibitem[\protect\citeauthoryear{Tang, Jiang, Li, Zeng, and Li}{Tang
  et~al\mbox{.}}{2015}]%
        {tang2015personalized}
\bibfield{author}{\bibinfo{person}{Liang Tang}, \bibinfo{person}{Yexi Jiang},
  \bibinfo{person}{Lei Li}, \bibinfo{person}{Chunqiu Zeng}, {and}
  \bibinfo{person}{Tao Li}.} \bibinfo{year}{2015}\natexlab{}.
\newblock \showarticletitle{Personalized recommendation via parameter-free
  contextual bandits}. In \bibinfo{booktitle}{\emph{Proceedings of the 38th
  International ACM SIGIR Conference on Research and Development in Information
  Retrieval}}. ACM, \bibinfo{pages}{323--332}.
\newblock


\bibitem[\protect\citeauthoryear{Tang, Rosales, Singh, and Agarwal}{Tang
  et~al\mbox{.}}{2013}]%
        {tang2013automatic}
\bibfield{author}{\bibinfo{person}{Liang Tang}, \bibinfo{person}{Romer
  Rosales}, \bibinfo{person}{Ajit Singh}, {and} \bibinfo{person}{Deepak
  Agarwal}.} \bibinfo{year}{2013}\natexlab{}.
\newblock \showarticletitle{Automatic ad format selection via contextual
  bandits}. In \bibinfo{booktitle}{\emph{Proceedings of the 22nd ACM
  international conference on Information \& Knowledge Management}}. ACM,
  \bibinfo{pages}{1587--1594}.
\newblock


\bibitem[\protect\citeauthoryear{Tewari and A.~Murphy}{Tewari and
  A.~Murphy}{2017}]%
        {Tewari2017}
\bibfield{author}{\bibinfo{person}{Ambuj Tewari} {and} \bibinfo{person}{Susan
  A.~Murphy}.} \bibinfo{year}{2017}\natexlab{}.
\newblock \showarticletitle{From Ads to Interventions: Contextual Bandits in
  Mobile Health}.
\newblock \bibinfo{journal}{\emph{Mobile Health: Sensors, Analytic Methods, and
  Applications}} (\bibinfo{date}{07} \bibinfo{year}{2017}),
  \bibinfo{pages}{495--517}.
\newblock
\showISBNx{978-3-319-51393-5}
\urldef\tempurl%
\url{https://doi.org/10.1007/978-3-319-51394-2_25}
\showDOI{\tempurl}


\bibitem[\protect\citeauthoryear{Yu, Lyu, and King}{Yu et~al\mbox{.}}{2017}]%
        {Yu2017CBRAPCB}
\bibfield{author}{\bibinfo{person}{Xiaotian Yu}, \bibinfo{person}{Michael~R.
  Lyu}, {and} \bibinfo{person}{Irwin King}.} \bibinfo{year}{2017}\natexlab{}.
\newblock \showarticletitle{CBRAP: Contextual Bandits with RAndom Projection}.
  In \bibinfo{booktitle}{\emph{AAAI}}.
\newblock


\end{thebibliography}
\end{document}